\documentclass[floatfix,twocolumn,showpacs,preprintnumbers,amsmath,amssymb,superscriptaddress,nofootinbib]{revtex4}
\usepackage{graphicx,psfrag} 
\newcommand{\newc}{\newcommand}
\newc{\gsim}{\lower.7ex\hbox{$\;\stackrel{\textstyle>}{\sim}\;$}}
\newc{\lsim}{\lower.7ex\hbox{$\;\stackrel{\textstyle<}{\sim}\;$}}
\newc{\gev}{\,{\rm GeV}}
\newc{\mev}{\,{\rm MeV}}
\newc{\ev}{\,{\rm eV}}
\newc{\kev}{\,{\rm keV}}
\newc{\tev}{\,{\rm TeV}}

\newc{\mz}{M_Z}
\newc{\mpl}{M_*}
\newc{\mw}{m_{\rm weak}}
\newc{\nr}[1]{N^c_R{}_{#1}}
\usepackage{amsmath}
\def\beq{\begin{equation}}
\def\eeq{\end{equation}}
\def\bea{\begin{eqnarray}}
\def\eea{\end{eqnarray}}
\def\bitem{\begin{itemize}}
\def\eitem{\end{itemize}}
\newc{\ie}{{\it i.e.}}          \newc{\etal}{{\it et al.}}
\newc{\eg}{{\it e.g.}}          \newc{\etc}{{\it etc.}}
\newc{\cf}{{\it c.f.}}
\def\bar#1{\overline{#1}}

\def\bra#1{\left\langle #1\right|}
\def\ket#1{\left| #1\right\rangle}
\def\abs#1{\left| #1\right|}

\def\inv{^{\raise.15ex\hbox{${\scriptscriptstyle -}$}\kern-.05em 1}}
\def\lbar{{\lower.35ex\hbox{$\mathchar'26$}\mkern-10mu\lambda}} 

\def\to{\rightarrow}

\let\<=\langle
\let\>=\rangle

\let\+=\uparrow
\let\al=\alpha
\let\be=\beta
\let\ga=\gamma
\let\Ga=\Gamma
\let\de=\delta

\let\ep=\epsilon

\let\la=\lambda

\let\si=\sigma

\let\th=\theta

\let\Om=\Omega

\let\al=\alpha
\let\be=\beta
\let\ga=\gamma
\let\Ga=\Gamma
\let\de=\delta

\let\ep=\epsilon

\let\la=\lambda

\let\si=\sigma

\let\th=\theta

\let\Om=\Omega

\begin{document}   
\title{WIMPonium and Boost Factors for Indirect Dark Matter Detection\footnote{Preliminary versions of this work were presented at PLANCK08, May 2008, by JMR, and at Southampton, Nottingham, Cambridge and Oxford, Feb-Oct 2008 by SMW.
}} 
\date{April 26th 2010.}
\author{John March-Russell}  
\email{jmr@thphys.ox.ac.uk}  
\affiliation{Rudolf Peierls Centre for Theoretical Physics, University of Oxford,
1 Keble Rd., Oxford OX1 3NP, UK}
\author{Stephen M. West}
\email{s.west1@physics.ox.ac.uk}
\affiliation{Rudolf Peierls Centre for Theoretical Physics, University of Oxford,
1 Keble Rd., Oxford OX1 3NP, UK}
\affiliation{Magdalen College, Oxford, OX1 4AU, UK}\begin{abstract}
We argue that WIMP dark matter can annihilate via long-lived ``WIMPonium" bound states in reasonable particle physics models of dark matter (DM).   WIMPonium bound states can occur at or near threshold leading to substantial enhancements in the DM annihilation rate, closely related to the Sommerfeld effect.  Large ``boost factor" amplifications in the annihilation rate can thus occur without large density enhancements, possibly preferring colder less dense objects such as dwarf galaxies as locations for indirect DM searches.  The radiative capture to and transitions among the WIMPonium states generically lead to a rich energy spectrum of annihilation products, with many distinct lines possible in the case of 2-body decays to $\gamma\gamma$ or $\gamma Z$ final states.   The existence of multiple radiative capture modes further enhances the total annihilation rate, and the detection of the lines would give direct over-determined information on the nature and self-interactions of the DM particles.

\end{abstract}   
\preprint{OUTP-08-01P}
\maketitle 
\section{\label{intro}Introduction}

One of the most promising ways in which to probe the nature of weakly-interacting-massive-particle (WIMP) dark matter is provided by indirect detection experiments in which the annihilation products of dark matter in astrophysical contexts are observed. Indirect detection experiments such as AMS \cite{ams}, ATIC \cite{atic}, EGRET \cite{egret}, GLAST \cite{glast},  HEAT \cite{heat}, HESS \cite{hess}, INTEGRAL \cite{integral}, PAMELA\cite{pamela}, and VERITAS \cite{veritas} look for signals ranging across energetic gamma rays, positron excesses, and anti-proton fluxes, and hints of deviations from background expectations now abound.  A common feature in the interpretation of these experiments is that the WIMP annihilation proceeds via simple, almost free-particle annihilation leading to a rate that depends on the WIMP relative velocity in only a very simple, essentially structureless way, and moreover, is directly related to the cross-section that led to the DM density at freeze-out. This assumption then implies that the primary astrophysical quantity determining the fluxes from DM annihilation is the local DM squared-density, $\rho({\bf x})^2$, with the velocity distribution of the DM being essentially irrelevant.  In addition the assumption of almost free-particle annihilation gives a relatively simple resulting energy spectrum for the annihilation products (though of course features such as steep falls at kinematic thresholds are generic).   
The dependence of the annihilation fluxes on $\rho({\bf x})^2$ has been widely employed in the suggested interpretations of the various experimental anomalies as, very frequently, a so-called ``boost-factor" in the annihilation rate is required to match the observed flux, and this is assumed to come from local over-densities of DM.

In this paper we point out that annihilation of WIMP dark matter via intermediate long-lived ``WIMPonium" bound states, $\Om^{(n,\ell)}$,
is possible in many particle physics models of DM \cite{ourtalks} (see \cite{posp} for another recent discussion of WIMPonium).  As we argue below, the WIMPonium bound states can occur at or near threshold in which case they lead to potentially very substantial (factors of $10^3$ to $10^5$)  enhancements in the DM annihilation rate, closely related to the well-known Sommerfeld non-perturbative enhancement \cite{sommerfeld} that has been applied to freeze-out calculations and indirect dark matter signals  \cite{freezeout,higgsportal,posp,others,jomass}.  The existence of this $\rho^2$- independent dynamically-induced ``boost factor" has important implications:  First large amplifications can occur in low-velocity-dispersion systems without a large (eg, cusp-like) density enhancement \footnote{Furthermore cusp-like dark matter distributions seem not to be favoured at present, see for example \cite{nocusp}.}, possibly preferring colder less dense objects such as dwarf galaxies as more promising places to search for DM signals compared to the higher density but higher velocity galactic center \footnote{See \cite{subir} for a travel guide to the directions in the sky where DM annihilation signals may be expected}.  Second, a knowledge of the full phase space density of the DM is necessary to reliably compute the DM annihilation rate, which strengthens further the case for detailed realistic simulations of the DM distribution in our galaxy.

In addition, the deeply-bound WIMPonium spectrum is often quite involved, and the radiative capture to and transitions among the various states generically lead to a rich energy spectrum of annihilation products, with many distinct lines (in the case of decays to $\gamma\gamma$ or $\gamma Z$ final states, similar to that found in WIMP annihilations \cite{bergetal}) possible. Although the observability, or otherwise of these distinct lines depends on the DM model being studied, and especially the resolution of the detector, the existence of all the various radiative capture modes further enhances the total annihilation rate, and the detection of even just a few of the lines would give direct information on the interactions of the DM particles.

\section{\label{BSbasics}Threshold Bound State Basics}

In order to demonstrate the important features of WIMPonium bound states it is useful to examine a simple, almost model independent, example. We introduce two complex scalar fields $s$ and $n$ with the following interactions and masses
\begin{gather} \label{simplag}
\mathcal{L}=\frac{\la^\prime \mu}{2} nss+\frac{m^2_n}{2}\abs{n}^2+\frac{m^2_s}{2}\abs{s}^2,
\end{gather}
where $\la^\prime$ is a dimensionless coupling and $\mu$, $m_n$ and $m_s$ are mass dimension one parameters.  We
assume that $\ep\equiv m_n/m_s \ll 1$ (typically we will take $m_n\sim m_Z$, while $500\gev \lsim m_s \lsim 30 \tev$).  A discrete symmetry is imposed forcing $s$ particles to appear in the Lagrangian in pairs so the $s$ scalar will be the stable dark matter WIMP.

As we have indicated in the introduction we are interested in two related phenomena:
1) The scattering of slow moving particles near bound state thresholds leading to amplification of the direct annihilation rate;  
2) The radiative capture to and decay of deeply bound states and transitions between different bound states. 

Since we are interested in low-velocity processes we can proceed by solving the Schr\"{o}dinger equation for the two-$s$-particle
system with the Yukawa potential $V=-\la^{2} e^{-m_nr} /8\pi r$ that follows from the exchange of $n$-scalars.  Here
$\la\equiv \la^{\prime}\mu/m_s$.  Semi-classical considerations show that the number of bound states $N_\ell$ of given orbital angular momentum $\ell$
satisfies \cite{bargmann} $(2\ell+1) N_\ell <  2 M_{r} \int r |V(r)| dr$, where $M_{r}$ is the reduced mass of the two
particle system, so, for example, the number of S-wave bound states in our case satisfies
$N_{0} <  \frac{m_s}{m_n}\al$, where $\al \equiv \la^{2}/8\pi$.
A more precise condition for there to be at least one bound state follows from numerical methods giving \cite{con} 
$\al \ge 0.84 m_n/m_{s}$.
Thus if we want to have a rich structure of energy levels our dark matter particles will either need to have large couplings with the $n$ scalar or need to have $m_{s}/m_n \gg 1$, or both.  We emphasize, however, that the most important phenomenology -- the large amplification of the DM annihilation rate -- requires only a single at-threshold bound state, and so imposes only a mild condition
on the coupling.  For instance for $m_n=m_Z$, $m_{s}=500\gev$, we require $\al\ge 0.15$
which is well within the perturbative regime $\al \lsim 2\pi$.   Note that, upon writing the complex scalar $s$ in terms of
its CP-even and odd parts $s=\phi_s + i a_s$, we have, due to Bose symmetry, that the $\phi_s \phi_s$ and $a_s a_s$ bound states
can only possess even orbital angular momentum $\ell =0,2,...$, while the angular momentum of the $\phi_s a_s$ bound
states is unrestricted.

\begin{figure}
\hspace{2cm}
\includegraphics[width=10cm]{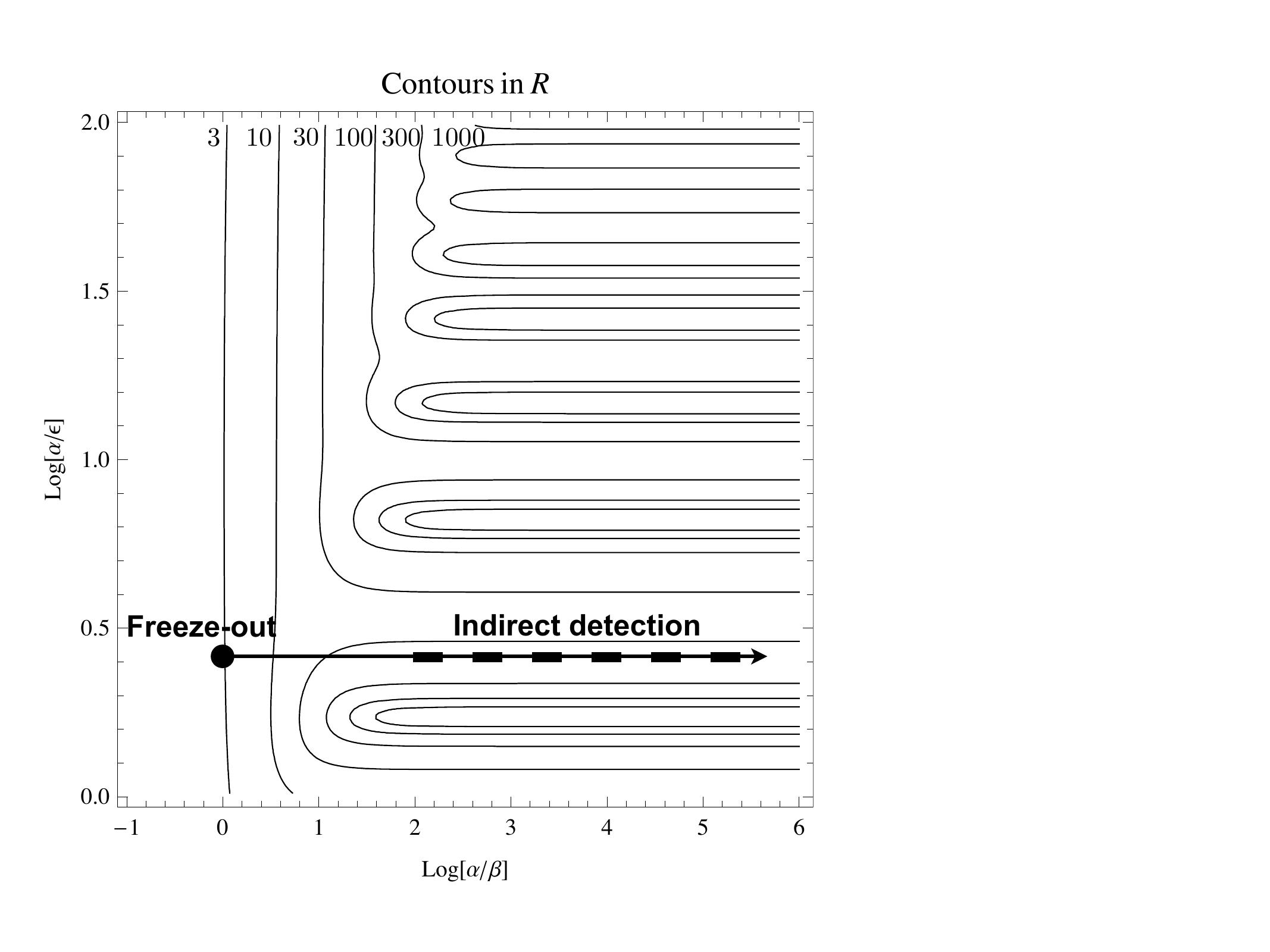}
\caption{\label{2d} Contour plot of the enhancement factor $R$.  Shown is a typical ``path" in parameter space as the velocity of the $s$ states decreases from the value at freeze-out to that relevant for indirect detection.
The path is not exactly horizontal due to thermal corrections to the masses and couplings.}
\end{figure}

\begin{figure}
\includegraphics[width=8.5cm]{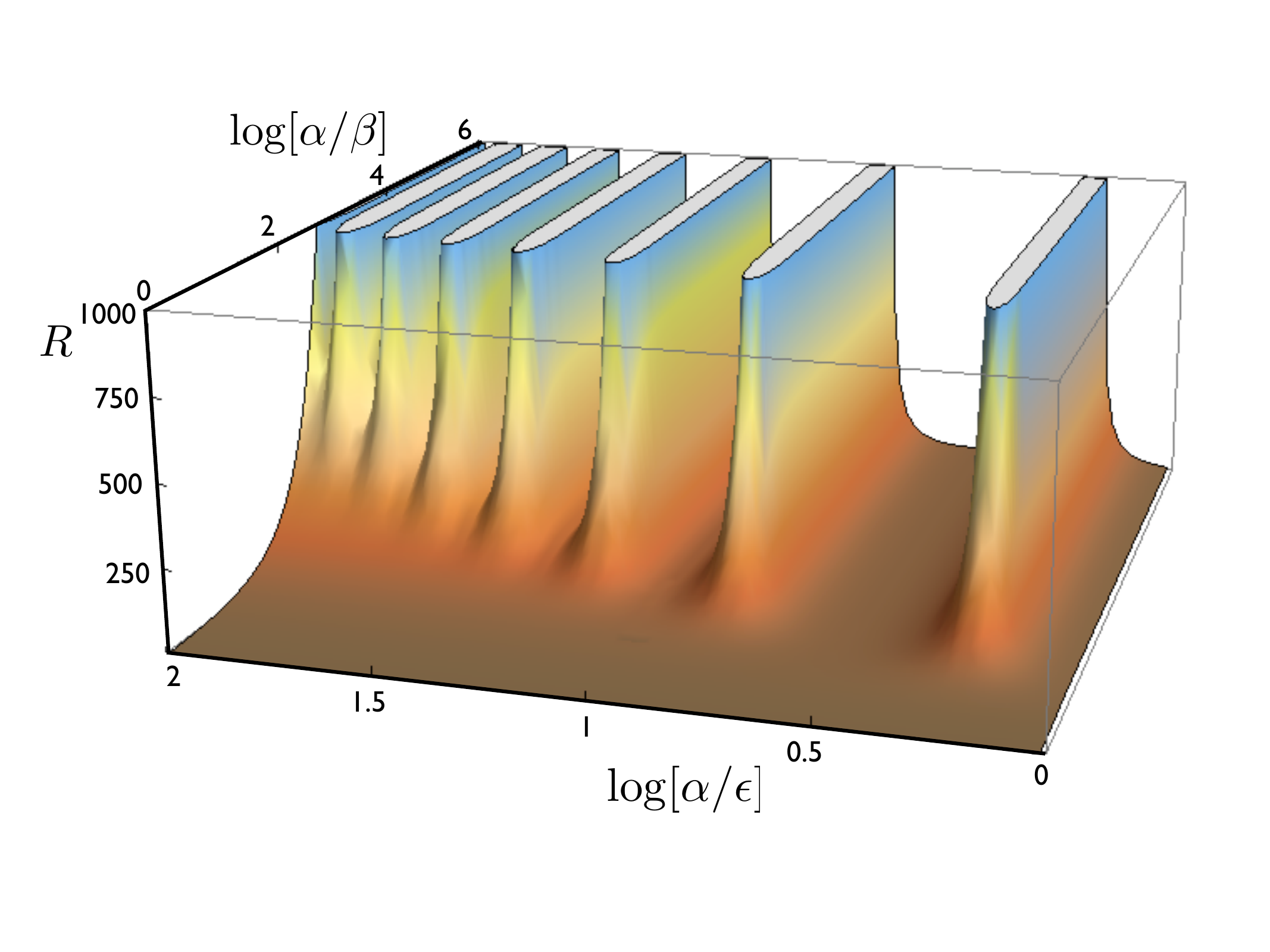}
\caption{\label{3d}A 3d version of Figure \ref{2d}.}
\end{figure}

Turning to the scattering of two slow moving $s$ particles by the Yukawa potential arising from the $n$ scalar exchange interaction, both elastic and inelastic scattering cross sections (such as radiative capture) are amplified by a non-perturbative Sommerfeld-like enhancement.  This enhancement can be formulated in terms of a non-relativistic quantum two-body problem with a potential acting between the incoming particles. To a good approximation this leads to a dressing of the dominant $S$-wave part of the tree-level cross sections by a multiplicative factor, 
\beq
\si^{l=0}=R\si^{l=0}_{tree}.
\label{rfactor}
\eeq
An exact analytic calculation of $R$ for a Yukawa potential is not possible (although we give a close analytic approximation later) and so we must proceed numerically. The Schr\"{o}dinger equation for the radial part of the two dark matter particle state, $\psi(r)$, with $l=0$, reads $-\psi^{\prime\prime}(r)/m_s+V(r)\psi(r)=E\psi(r)$,  where $E=m_s\be^2$ is the kinetic energy of the two dark matter particles in the center-of-mass frame, where each dark matter particle has velocity $\be$. 
Using the outgoing boundary conditions, $\psi^{\prime}(\infty)/\psi(\infty)=im_s \be$, $R$ is given as $R=\abs{\psi(0)/\psi(\infty)}^2$. Rewriting $r=y/m_n$ and letting $\ep=m_n/m_s$ we can rewrite the 
Schr\"{o}dinger equation as a function of the two ratios $\al/\ep$ and $\al/\be$, {\it viz}
\begin{gather}
-\left(\frac{d^2}{dr^2}+\frac{\al}{\ep}\frac{e^{-y}}{y}\right)\psi(y)=\frac{\be^2}{\al^2}\frac{\al^2}{\ep^2}\psi(y).
\label{schro}
\end{gather}
The resulting numerical solutions for $R$ are functions of $t\equiv \al/\ep$ and $u\equiv\al/\be$ with the 2d contour plot shown in Fig \ref{2d} and the 3d version in Fig \ref{3d}. 

It is clear that there are two distinct regions in Fig \ref{2d}. For large values of the velocity ($\be\gsim \ep$) there is a relatively flat region -- the Coulomb region.  This is the part of parameter space relevant for freeze-out, the magnitude of the enhancements being at most a factor of 3  to 5 \cite{freezeout,higgsportal}. More interesting is the low-velocity region in which we see the effect of resonance peaks. These peaks correspond to the formation of $l=0$ bound states at threshold ($E=0$), and are the focus of this analysis.

Provided we are sufficiently close to a resonance peak, the dependence of $R$ on $t=\al/\ep$ and $u=\al/\be$ is described by a modified Breit-Wigner resonance formula applicable for threshold resonances due to Bethe and Placzek (BP) \cite{betheplaczek,landl}. Considering only elastic scattering, the Breit-Wigner resonance cross section is
\begin{gather}
\si_e^{BW}=\frac{\pi}{k^2}\frac{\Gamma^2_e}{((m_s\be^2-\varepsilon_0)^2+\Gamma^2/4)},
\end{gather}
where $\Gamma_e$ and $\Gamma$ are the elastic dissociation and total width for the resonant bound state and $\varepsilon_0$ is the distance of the resonance from exact zero-energy and is independent of $\be$.
Following BP, for small $\be$ near a threshold resonance the BW expression is modified
by replacing $\Gamma_e=\sqrt{E}\gamma_e$ with $\gamma_e$ not depending on $E$ . Using $E=k^2/m_s$, we  have
\begin{gather}
\si_e^{BP}=\frac{\pi}{m_s}\frac{\gamma^2_e}{((m_s\be^2-\varepsilon_0)^2+m_s\be^2\gamma_e^2/4)}
\label{bpcr} 
\end{gather}
(as $R$ has been found by including only elastic scattering we have set $\Gamma=\Gamma_e$).

\begin{figure}
\includegraphics[width=10cm]{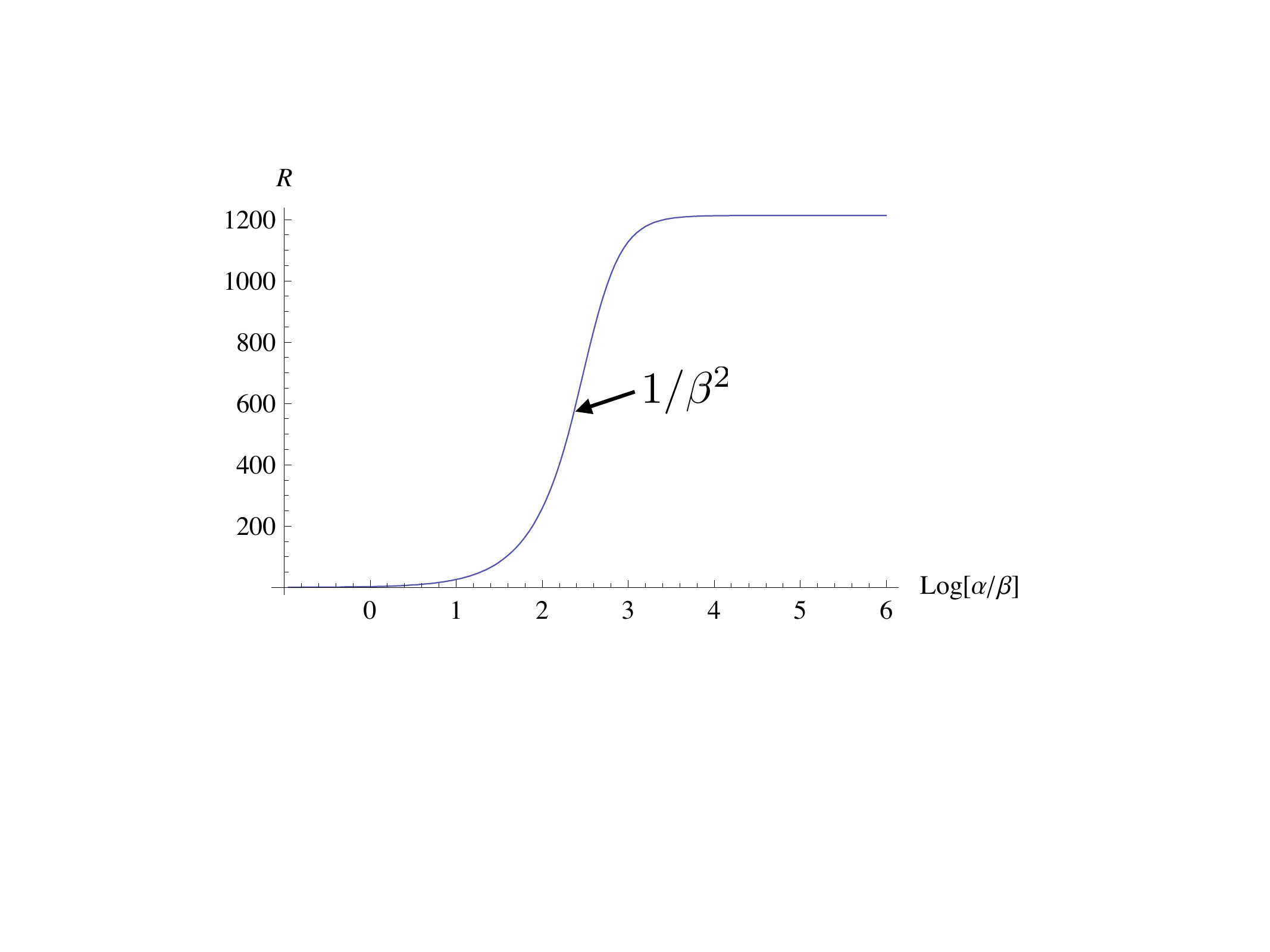}
\vspace{-2cm}
\caption{\label{sliceep}The $\be$-dependence of $R$ at a fixed value of $\al/\ep$ as derived from the numerical solution of Eq.(\ref{schro}).}
\end{figure}

Eq.(\ref{bpcr}) shows that for $\varepsilon_0\neq 0$ the cross-section first increases as $1/\be^2$ and then becomes independent of $\be$ for $\be\ll 1$ as shown in Fig \ref{sliceep}.  The plateau begins at $\be\sim\sqrt{|\varepsilon_0|/m_s}$, and corresponds to $\si_e^{BP}\simeq 4\pi/(m_s |\varepsilon_0|)$.  

Comparing the numerical calculation of the elastic cross section to $\si_e^{BP}$ in the low-velocity limit
we can extract the  numerical values of $\gamma_e$ and $\varepsilon_0$ as a function of the parameters
$\al$ and $\ep$ for each of the possible near-threshold bound state resonances. 

The plateau arises as $\be\rightarrow0$ (for $\varepsilon_0\neq 0$) with asymptotic value of the cross section
\beq
\si_e^{BP}|_{\be\rightarrow0}=\frac{\pi}{m_s}\left(\frac{\gamma_e}{\varepsilon_0}\right)^2 .
\eeq
Since $\si |_{\be\rightarrow0}=\bar{R}(t)\si^{l=0}_{tree}$,
where $\bar{R}(t)$ is independent of $\be$, 
\beq
\gamma_e^2=\varepsilon_0^2\frac{m_s}{\pi}\bar{R}(t)\si_{tree}.
\label{game}
\eeq
The function $\bar{R}(t)$ has the form $\bar{R}(t)=a^{(n l)}/(t-t_0^{(n l)})^2$, where $a^{(n l)}$ and $t_0^{(n l)}$
depend upon the principal and orbital angular momentum quantum numbers $(n l)$ of the resonance that
is close to threshold as $t$ is varied.  Table \ref{fits} gives numerical fits for the $S$-wave resonances. 

\begin{table}\centering
\begin{tabular}{|c|c|c|}
\cline{1-3}
\vbox to1.93ex{\vspace{1pt}\vfil\hbox to11ex{\hfil Resonance\hfil}} & 
\vbox to1.93ex{\vspace{1pt}\vfil\hbox to8ex{\hfil $a$\hfil}} & 
\vbox to1.93ex{\vspace{1pt}\vfil\hbox to8ex{\hfil $t_0$\hfil}} \\

\cline{1-3}
\vbox to1.63ex{\vspace{1pt}\vfil\hbox to11ex{\hfil $1s$\hfil}} & 
\vbox to1.63ex{\vspace{1pt}\vfil\hbox to8ex{\hfil 12.2\hfil}} &  
\vbox to1.63ex{\vspace{1pt}\vfil\hbox to8ex{\hfil 1.681\hfil}} \\

\cline{1-3}
\vbox to1.63ex{\vspace{1pt}\vfil\hbox to11ex{\hfil $2s$\hfil}} & 
\vbox to1.63ex{\vspace{1pt}\vfil\hbox to8ex{\hfil 198.5\hfil}} & 
\vbox to1.63ex{\vspace{1pt}\vfil\hbox to8ex{\hfil 6.453\hfil}} \\

\cline{1-3}
\vbox to1.63ex{\vspace{1pt}\vfil\hbox to11ex{\hfil $3s$\hfil}} & 
\vbox to1.63ex{\vspace{1pt}\vfil\hbox to8ex{\hfil 1015.9\hfil}} & 
\vbox to1.63ex{\vspace{1pt}\vfil\hbox to8ex{\hfil 14.358\hfil}} \\

\cline{1-3}
\vbox to1.63ex{\vspace{1pt}\vfil\hbox to11ex{\hfil $4s$\hfil}} & 
\vbox to1.63ex{\vspace{1pt}\vfil\hbox to8ex{\hfil 3220.6\hfil}} & 
\vbox to1.63ex{\vspace{1pt}\vfil\hbox to8ex{\hfil 25.407\hfil}} \\

\cline{1-3}
\vbox to1.63ex{\vspace{1pt}\vfil\hbox to11ex{\hfil $5s$\hfil}} & 
\vbox to1.63ex{\vspace{1pt}\vfil\hbox to8ex{\hfil 7849.0\hfil}} & 
\vbox to1.63ex{\vspace{1pt}\vfil\hbox to8ex{\hfil 39.609\hfil}} \\
\cline{1-3}
\end{tabular}
\caption{Numerical fits of $\bar{R}(t)=a/(t-t_0)^2$. The cross sections are extremely sensitive to the values of $t_0$.   The values given have been rounded off so that they can be displayed.\label{fits}}
\end{table}

We remark in passing that an exact analytic treatment is possible for the Hulth\'{e}n potential $V_H(r)=-\al m_ n e^{-rm_n}/(1-e^{-rm_n})$ which has similar $r\rightarrow 0$ and $r\rightarrow \infty$ behaviour to the Yukawa potential.  The S-wave phase shifts are (see also \cite{hulthepot}) 
\begin{align}
\de_0=\frac{\pi}{2}-\mbox{arg}\left[\Ga\left(1+\frac{ik}{m_n}-\sqrt{A-\frac{k^2}{m^2_n}}\right)\right]+~~~\\
\mbox{arg}\left[\Ga\left(\frac{2ik}{m_n}\right)\right]-\mbox{arg}\left[\Ga\left(1+\frac{ik}{m_n}+\sqrt{A-\frac{k^2}{m^2_n}}\right)\right]\nonumber,
\end{align}
where $k=m_s\be$ is the momentum of the scattered state and $A$ can be thought of as $t=\al/\ep$ up to a factor of two. From this the S-wave elastic cross section $\si_e=4\pi\abs{f}^2$ follows using $f=\left(e^{2i\de_0}-1\right)/2ik$.
Taking the $\be\rightarrow 0$ limit of $\si_e$, the dominant behaviour is
$\si \propto 1/(A-A_0)^2$, where $A_0$ plays the same role as $t_0^{(n0)}$.  Exactly on one of the resonances, $A= A_0$, the dependence on $\be$ is $\si_e \propto 1/\be^2$ which agrees with the BP form.

So far we have only discussed the case of elastic scattering. However, as we are interested in the indirect signals coming from dark matter annihilations we need to examine the case of inelastic scattering. In order to write down the inelastic cross section in the presence of the long range enhancements we again follow BP by writing 
\begin{gather}
\si_{i}^{BP}=\frac{\pi}{k^2}\frac{\Gamma_e\Gamma_i}{((E-\varepsilon_0)^2+\Gamma^2/4)},
\label{ine1}
\end{gather}
where $\Gamma_i$ is the inelastic width associated with the direct annihilation or radiative capture of the incoming $s$-pair. This form is again only applicable when we are suitably close to one of the S-wave resonances. Following BP the inelastic width is a constant as opposed to the energy dependent elastic width. Substituting the form for the elastic width $\Gamma_e=\sqrt{E}\ga_e$ into Eq.(\ref{ine1}) we have
\begin{gather}
\si_{i}^{BP}=\frac{\pi}{\be m_s^{3/2}}\frac{\ga_e\Gamma_i}{[(m_s\be^2-\varepsilon_0)^2+\frac{(\Ga_i+\be \sqrt{m_s}\ga_e)^2}{4}]}. 
\label{ine2}
\end{gather}

The first point to note here is the $1/\be$ enhancement of the inelastic cross section compared to the elastic -- the usual Bethe $1/v$ law \cite{bethe}. This follows if we take the $\be\rightarrow 0$ limit of Eq.(\ref{ine2}) (assuming that we are not exactly on resonance)
\begin{gather}
\si_{i}^{BP}=\frac{\pi}{\be m_s^{3/2}}\frac{\ga_e\Gamma_i}{(\varepsilon_0^2+\Ga_i^2/4)},
\label{ine3}
\end{gather}
compared to the $\be\rightarrow 0$ elastic cross section,
$\si_{e}^{BP}=\pi\ga^2_e/ m_s(\varepsilon_0^2+\Ga_i^2/4)$.  Second, for $m_s\be^2> \varepsilon_0, \Ga_i$, the inelastic cross section Eq.(\ref{ine2}) shows a $1/\be^3$ dependence which plateaus to the standard $1/\be$ dependence when  $m_s\be^2\ll \varepsilon_0, \Ga_i$. This behaviour is exactly that of Eq.(\ref{rfactor}) for the enhancement of the na\"{i}ve inelastic cross section given the $\be$ dependence of $R$ plotted in 
Fig. \ref{sliceep} resulting from numerical solutions and as discussed above.  Third, the final limiting value of $\be\si_i^{BP}$ has a  $1/\varepsilon_0^2$ dependence and shows that for WIMPonium bound states close to threshold the cross sections are further enhanced. (In the above we have made the assumption that $\varepsilon_0> \Ga_i$ which is true over the majority of parameter space. However, for the case where $\varepsilon_0< \Ga_i$ the true dependences can be more complicated involving cross terms of the inelastic and elastic widths.) Depending on the size of $\varepsilon_0$ the size of the $1/\be^2$ enhancements can be significant since for DM annihilations in present day astrophysical systems the relevant range of $\be$ is $\sim 10^{-3}  - 10^{-5}$.

A useful way of thinking about and calculating approximately the amplification of the elastic and inelastic cross sections is in terms of the diverging scattering lengths that occur when a bound state energy tends to zero.  Recall from the elementary
theory of non-relativistic elastic scattering that the S-wave phase shift satisfies 
\beq
\lim_{k\rightarrow 0} k \cot[\de_0(k)] = -\frac{1}{l_s} + \frac{1}{2} r_e k^2 +...
\label{phaseshift}
\eeq
where $r_e\sim 1/m_n$ is the effective range of the potential and $l_s$ is the scattering length which is related
to the near-threshold bound state energy $\varepsilon_0$ by
\beq
\ell_s = \frac{1}{\sqrt{m_s |\varepsilon_0 | }} +....
\label{scatteringlength}
\eeq
These equations imply $\cot \de_0 = -\sqrt{| \varepsilon_0 |/E}$ where $E=k^2/m_s$ is the CM scattering energy, and we have assumed $k\to 0$
before $\varepsilon_0\to 0$.  From the standard expression of the elastic cross-section in terms of $\de_0$ one then finds agreement
with the BP formula Eq.(\ref{bpcr}) in the same limit of $k\to 0$
before $\varepsilon_0\to 0$.  Comparing with the BP form we learn that $\ga_e^2 = 4 |\varepsilon_0| \rightarrow 0$ as the bound state approaches exactly the zero-energy threshold.  The advantage of this approach is that the S-wave component of the distorted incoming plane waves can be simply expressed in terms of $\de_0$ and thus $\varepsilon_0$.  To a good approximation the wavefunction is
\beq
\psi_{l=0}\approx\frac{\sin(kr + \delta_0)}{kr}.
\label{spartialwave}
\eeq
This form and its dependence on the scattering length $\ell_s$ will be useful to us when we discuss
radiative capture.

\section{\label{decays} Decays, Captures, \& Transitions}

WIMPonium bound states possess a rich phenomenology of radiative captures to various bound states, and decays and transitions from
or among the bound states.

\subsection{\label{radiativecapture} Radiative Capture}

For interesting bound-state to bound-state transitions to be relevant, the radiative capture cross section must be significant.   Similar to elastic scattering the radiative capture cross section is enhanced when there is a near-threshold bound state, as is approximately the situation in neutron-proton scattering where enhanced radiative capture to a bound deuteron is possible. 

We will consider transitions into both the $\ell=0$ and $\ell=1$ bound state energy levels. The most economical way to perform such a calculation is to first expand the continuum state in terms of partial waves where we will keep only the  S-partial wave, Eq.(\ref{spartialwave}) (the incoming P-wave gives terms that are suppressed by $\be^2$). 
Labeling the bound state wavefunctions into which we will be capturing as $\psi_{nll_z}$, the two most important states are $\psi_{100} \approx (\pi a_0^3)^{-1/2} \exp(-r/a_0)$ and $\psi_{210}\approx (32\pi a_0^3)^{-1/2} r \cos{\th} \exp(-r/2a_0)/a_0$, where we've assumed that the $\Om^{(n,\ell)}$ bound state wavefunctions are similar to hydrogen with $a_0\sim 2/m_s\al$.  (The bound states $\psi_{200}$ and $\psi_{21\pm1}$ do contribute, with the total rates for radiative capture changing by, at most, an $\cal{O}$(1) factor. As we are interested in the general parametric dependence we neglect the contributions from capture into these bound states.)  The radiative capture cross section depends on matrix elements of the form $I=\langle\psi_{f}|\mathcal{O}|\psi_{i}\rangle$, where $\psi_{i}(r)$ is the initial distorted partial wave of the continuum state, $\psi_{f}(r)$ is the wavefunction of the bound state into which we are being captured, and $\mathcal{O}$ is the interaction Hamiltonian. In our case $\mathcal{O}=\frac{\la}{2}e^{i{\bf p.r}}$, where $p$ is the momentum of the radiated scalar $n$ state.


Expanding the exponential in powers of ${\bf p.r}$ and considering radiative capture into the $1s$ and $2p$ states, the overlap integrals take the forms
\bea
I_{1s}^2&=&\frac{\la^2}{4}\left| \int \left[\psi_{100}({\bf p_1.r})^2\psi_{l=0}\right] d^3r \right|^2,\nonumber\\
I_{2p}^2&=&\frac{\la^2}{4}\left| \int \left[\psi_{210}({\bf p_2.r})\psi_{l=0}\right] d^3r \right|^2,
\eea
where the first non-zero integral comes at second order in ${\bf p.r}$ for capture into the $1s$ state and at linear order for capture into the $2p$ state. 


From this the rates are found to be 
\bea
\Ga_{1s}&\approx& 288\pi\, \al  m_s^3(a_0p_{1})^5(4a_0+\de_0/k)^2,\nonumber\\
\Ga_{2p}&\approx& 1024\pi\, \al  m_s^3(a_0p_{2})^3(8a_0+\de_0/k)^2,
\label{radrates}
\eea
where $p_1\approx\sqrt{E_1^2-m_n^2}$ and $p_2\approx\sqrt{E_2^2-m_n^2}$ are the momenta of the $n$ scalars due to transitions into the $1s$ and $2p$ states respectively. For simplicity, from now on we will assume that $m_n$ is small relative to transition energies and so can be neglected.  Clearly if $m_n$ is not small then there are trivial kinematical
suppression factors.

Using the S-wave phase shift, Eq.(\ref{phaseshift}), and taking the small $k$ limit gives $\de_0\approx k\ell_s$. For large scattering length (compared with $8a_0$) the rates become
\bea
\Ga_{1s}&\approx & 9\pi \al^6\, m_s^3  \ell_s^2 \approx 9\pi \al^6 \frac{m_s^2}{\varepsilon_0},\nonumber\\
\Ga_{2p}&\approx & 2\pi \al^4\, m_s^3  \ell_s^2 \approx 2\pi \al^4 \frac{m_s^2}{\varepsilon_0},
\eea
where in the last expression we have used the relation,  Eq.(\ref{scatteringlength}), between the scattering length $\ell_s$ and the near threshold bound state energy and continuing the analogy with hydrogen we have taken the bound state energies to be $E_n=-m_s\al^2/4n^2$. This implies radiative capture cross sections
\bea
\si^{RC}_{1s} &\approx&  \frac{9\pi\al^6}{\be \varepsilon_0 m_s},\\
\si^{RC}_{2p} &\approx&  \frac{2\pi\al^4}{\be \varepsilon_0 m_s}.
\label{radiativecapturecross}
\eea
This shows the usual $1/\be$ Bethe-dependence of an inelastic cross section near $\be\rightarrow 0$, and most importantly an {\it additional}
$\al^2 m_s/\varepsilon_0$ enhancement relative to the radiative capture cross section if there was not a near threshold bound state (in other words if the $a_0$- dependent terms dominated over $\de_0/k$ in Eq.(\ref{radrates})).  The factor $\al^2 m_s/\varepsilon_0$ is just the ratio of the typical bound state energy compared to the near-threshold energy. In addition to this, it should be noted that for $\al\sim\cal{O}$(1) (which is still well within the perturbative regime) the radiative capture cross sections are further increased, although the simple hydrogen-like scaling that we have employed starts to break down and numerical methods must be used.

\subsection{Decays}

Consider annihilation of the $ss$ bound state $\Omega^{(n,\ell)}$ to light (${\rm mass} \ll m_s$) degrees of freedom.  Let the amplitude for the free $2\to 2$ scattering be $A_{fi}$, then
from the standard theory of, eg., positronium decay the amplitude $M_{fi}$ for the bound state decay is
\beq
M_{fi} = \frac{1}{4 m_s} \sqrt{\frac{M_\Omega}{\pi^3}} \int d^3p \, {\tilde \psi}(p) A_{fi}(p), 
\eeq
where the momentum-space bound state Schr\"{o}dinger wavefunction is
normalized as $\int d^3p |{\tilde \psi}(p)|^2 = 1$. Here $p$ is defined by $(0,2 {\bf p}) = q_1-q_2$ where $q_i$ are the 4-momenta of the two $s$ constituents.  In the limit of relatively weak binding
($E_B << m_s$) we have $M_\Omega \approx 2m_s$.  Expanding $A_{fi}$ in powers of $p^2$,
$A_{fi}(p) = A_{fi}^{(0)} + p^2 A_{fi}^{(2)} +...$ and using ${\tilde \psi}(p) = \int d^3 x e^{ip.x}\psi(x)/(2\pi)^{3/2}$ gives
\beq
M_{fi} \approx A_{fi}^{(0)} \psi(0) - \frac{15 A_{fi}^{(2)}}{2} \frac{\partial^2 \psi(0)}{\partial r^2} + ...,
\label{boundstatedecay}
\eeq
which enables the calculation of the decay width of various orbital angular momentum bound states.  

For $m_s/m_n \gg 1$, and the $ssn$ coupling $\la$ not very large, the wavefunctions of the bound states are approximately hydrogen-like, which upon applying
Eq.(\ref{boundstatedecay}) gives the annihilation width $\ell=0$
\beq
\Gamma_{ann}^{\ell=0} \simeq \frac{\alpha^{5}}{3 n^3} m_s .
\label{decayrate}
\eeq
This expression is most accurate for the more tightly bound WIMPonium states with small principal quantum number, $n=1,2,...$, while for higher bound states a precise decay width requires numerical evaluation of the bound state wavefunction.  Parametrically Eq.(\ref{decayrate}) gives a good estimate of the decay width in all cases.

From Eq.(\ref{boundstatedecay}) and the expansion of the free $2\to 2$ scattering amplitude, the decay widths of the higher orbital angular momentum states are parametrically suppressed 
by powers of $1/(m_s a_0)^2$.  Using hydrogen-like scaling once again shows that, eg,  $\Gamma_{ann}^{\ell=1}/\Gamma_{ann}^{\ell=0}
\simeq \alpha^2$ which, if we take $\al<1$, is parametrically small. However, if we take $\al\sim\cal{O}$(1) all decay widths can be large and comparable.

\subsection{\label{transitions} Transitions}

Transitions between the various bound states are possible with either the emission of 
the $CP=\pm 1$ components of $n$, or if on-shell $n$-production is kinematically disallowed,
by decay to light SM states through virtual $n$-emission.    First
assuming on-shell $n$ emission (possible when $\al^2 m_s \gsim m_n$), the relevant WIMPonium matrix element reduces to $T_{fi} = \bra{R_{n' \ell'} Y_{\ell' m'}} \la \exp(i{\bf pr}) \ket{R_{n \ell} Y_{\ell m}}$.
Similar to transitions in hydrogen-like systems we may expand the exponential in powers $kr\sim \al$ (for relativistic $n$).  The first transitions occur at ${\cal O}(kr)$ with the emission of the $CP=-1$ state of $n$ with orbital angular momentum $l=1$, changing the bound state from
$\phi_s\phi_s$ or $a_s a_s$ to $\phi_s a_s$ or vice versa.   

A good estimate of the various transition rates follows from a straightforward application of Fermi's Golden Rule.  Parametrically the rate for $\Delta \ell=1$ transitions emitting a relativistic 
$a_n$ scales as
\beq
\Gamma_{\rm trans}^{\Delta \ell=1} \simeq \al a_0^2 \, (\Delta E)^3 \sim \al^5 m_s
\label{transitionrate}
\eeq
where $\Delta E$ ($\gg m_n$) is the energy splitting between the bound states.   We see the $(\Delta E)^3$ behaviour familiar from hydrogen-like systems which favours deep transitions.  
All other transitions emitting an on-shell $n$ are down by powers of $(\Delta E a_0)^2 \sim \al^2$.  For instance $\Delta \ell = 0$ or $2$ transitions emitting a $CP=+1$ $\phi_n$ state are suppressed as $\Gamma_{\rm trans}^{\Delta \ell=0,2}/\Gamma_{\rm trans}^{\Delta \ell=1} \sim \al^2$ as well as by final state phase space factors $(\Delta E/\Delta E')^3$.  Transitions to light SM states via virtual $n$'s are further suppressed by both couplings such as $\al$ or $\alpha_{em}$ and (at least) three-body phase space factors.   

\begin{figure}
\includegraphics[width=8.5cm]{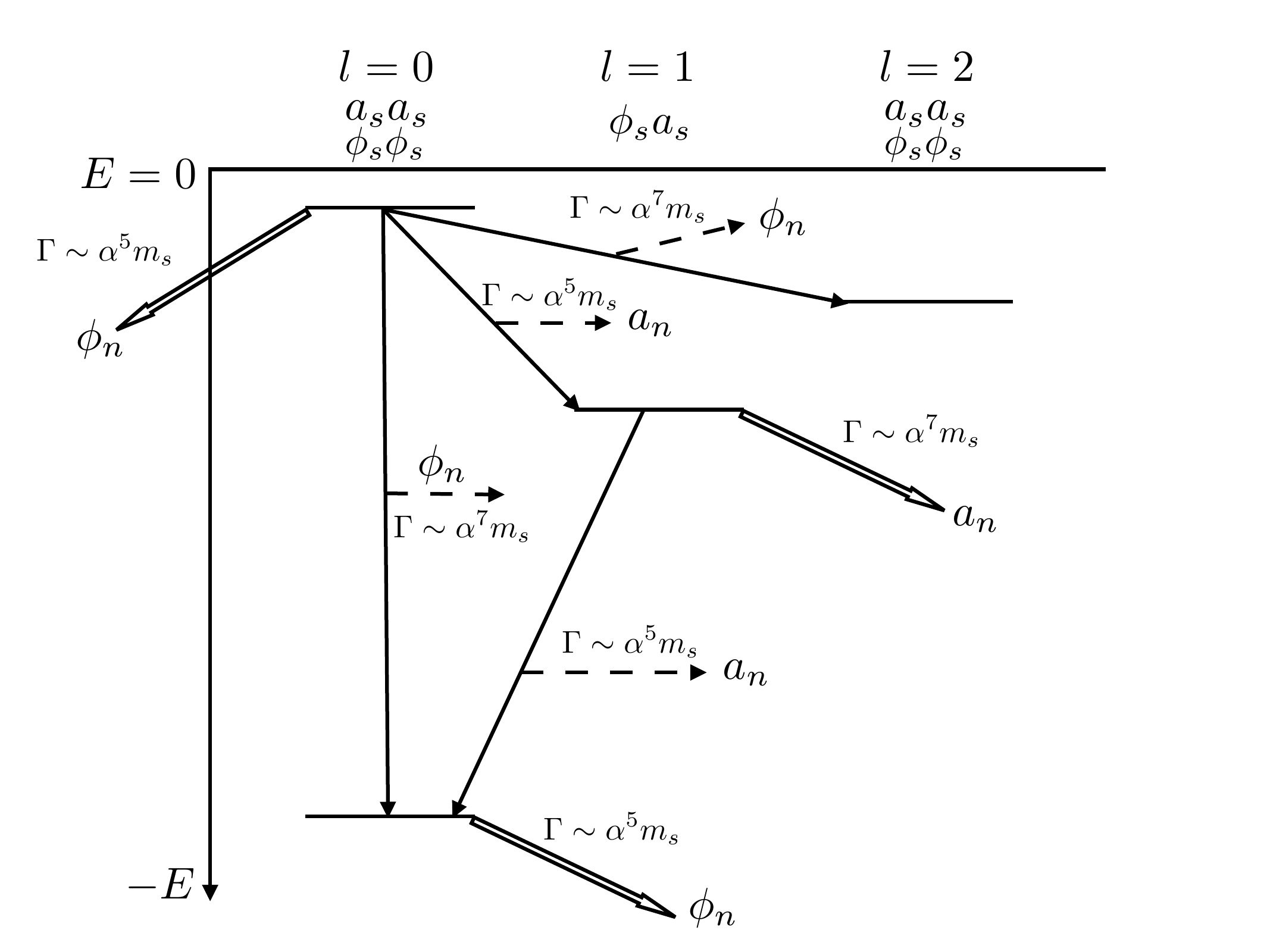}
\caption{\label{energylevel} Diagram illustrating, for various bound states, the rates of annihilation (double line arrows), and transitions emitting a $\phi_n$ or $a_n$ (single line arrows). We have omitted the $\phi_sa_s$ $l=0$ and $l=2$ bound state energy levels and the associated transitions and decays for clarity.}
\end{figure}

\begin{figure}
\includegraphics[width=8.5cm]{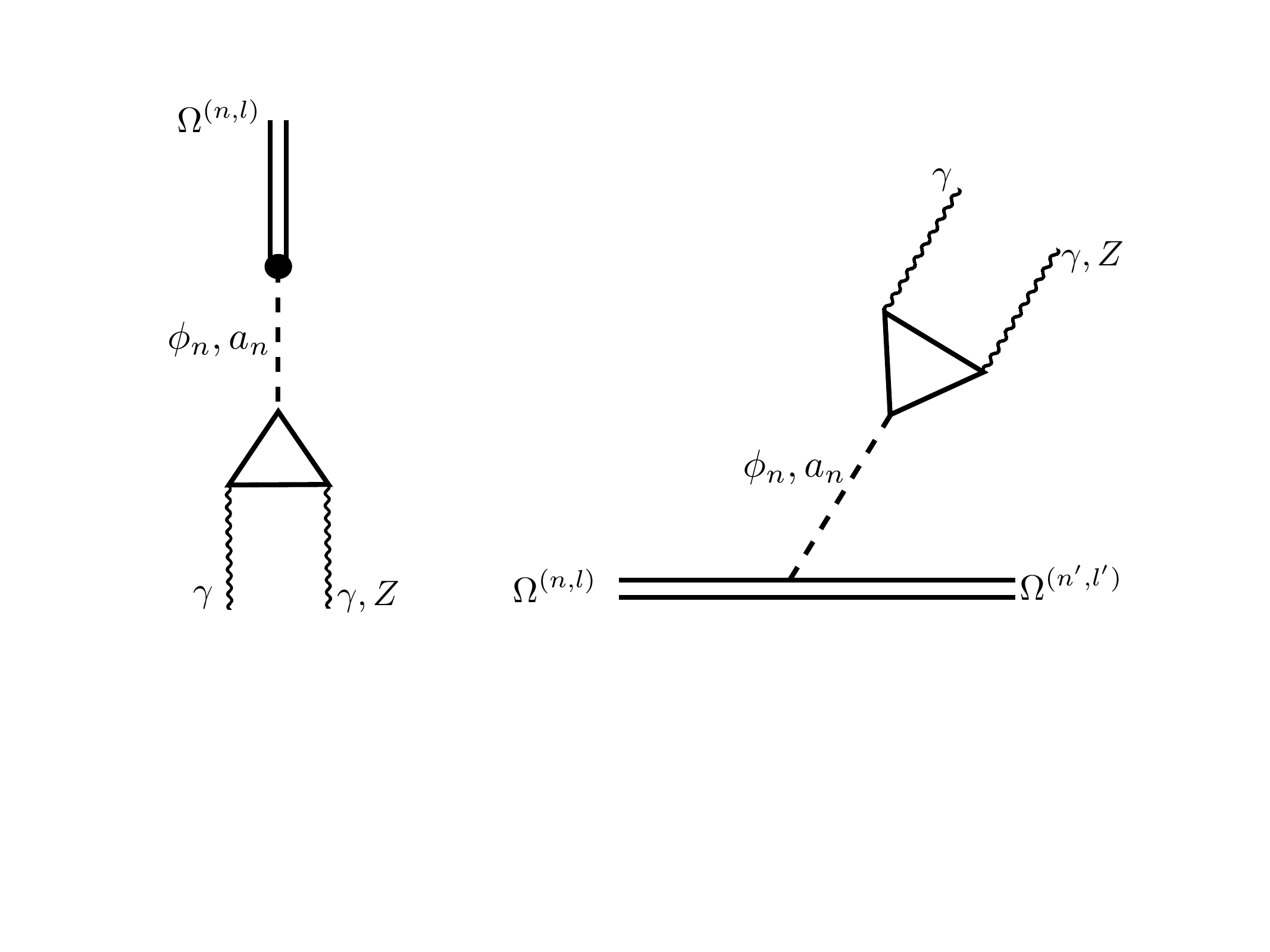}
\vspace{-2cm}
\caption{\label{photonemission}Schematic diagrams showing annihilation of, or transition between, different bound states leading to discrete $\ga\ga$ or $\ga Z$ lines.  The states in the triangle can be any charged scalar, $W^\pm$ gauge boson, or fermion that couples to $\phi_n, a_n$, in general via mixing with Higgs states.}
\end{figure}

The most important feature of Eq.(\ref{transitionrate}) is that it shows that the transition rate
between different bound states can be competitive to the direct decay rate, Eq.(\ref{decayrate}),
of the $\ell=0$ tower as long as on-shell $n$ production is kinematically possible. If the $s$ DM particles are captured in a P-wave orbital angular momentum state, then the extra
suppression of P-wave annihilations implies that transitions via on-shall $n$'s can dominate.  Furthermore transitions to light SM states via virtual $n$'s might be only mildly suppressed 
relative to P-wave annihilations depending on the strength of the coupling of $n$ to Higgs (and thus other SM) fields. Fig.\ref{energylevel} schematically illustrates the dominant decays and transitions and their respective rates while Fig.\ref{photonemission} depicts diagrams responsible for discrete $\ga$ lines.

\section{\label{consequences} Consequences for Indirect Detection and LHC}

A wide range of consequences follow from the existence of WIMPonium bound states.  As we have already mentioned
they provide a new, dynamical mechanism for the ``boost factors'' that are often introduced to explain anomalies in indirect
detection observations.  Unlike traditional $\rho(x)^2$ enhancements they depend on the velocity distribution
of the DM particles, and since between the freeze-out epoch and today the velocity changes from $\be_{fo} \sim 1/5$ to
$\be_{now}\sim 10^{-3}-10^{-5}$, the near-threshold bound state enhancement decouples the value of the annihilation
cross section determined by successful thermal freeze-out from that observed today in indirect annihilation observations. (See
\cite{higgsportal} for a discussion of thermal freeze-out production of DM in models closely related to our toy theory.)
In fact, because of thermal corrections to the couplings and masses during freeze-out the ``path" taken in $(\al/\be, \al/\ep)$
parameter space as the universe cools is not exactly a $\al/\ep=$constant line, but instead can move towards or away from the
value at which a resonance occurs exactly at threshold.  This further decouples the value of the cross-section at freeze-out from
that observed now.  Moreover, the enhancement of the cross-section as the universe cools leads to the possibility
that there is interesting residual post-freeze-out annihilation of the DM particles, for instance leading to changes
in BBN predictions of such elements as $^6$Li and $^7$Li \cite{workinprogress}. Depending on the maximum size of the
threshold enhancement, which is determined both by the size of the inelastic widths, and the degree to which the
bound state approaches zero energy, this dynamical boost factor can be large enough to potentially favour
environments such as dwarf galaxies, since their velocity dispersions go down as
low as 3-5${\rm km s}^{-1}$ \cite{dwarfgalaxies} compared to the typical galactic value of $\sim 200-300{\rm km s}^{-1}$.

Turning to particle physics model-building issues, the existence of WIMPonium bound states implies that the standard supersymmetric neutralino DM picture must be modified somewhat.  Although we have explained the phenomenology of WIMPonium in the context of a very simple, purely scalar toy model, we emphasize that similar phenomena are possible if the DM is fermionic, or even neutralino DM.
From the condition for the existence of at least one bound state $\al \ge 0.84 m_n/m_s$ we learn that if the DM is a neutralino 
interacting via the exchange of $W^\pm$ and $Z$ gauge bosons (and Higgs states) then the neutralino must be heavy
$m_{\rm neutralino} \gsim m_W/\al_2 \sim 2\tev$.  On the other hand if the DM particle interacts with a new strong-interacting sector, say a hidden valley sector\cite{hiddenvalley}, then the DM particle can potentially have close-to-weak-scale mass.

 A particularly attractive possibility is to have the DM state associated with electro-weak symmetry breaking dynamics in some way, so that it interacts with
the Standard Model via the so-called Higgs Portal \cite{higgsportal}, and has strong Higgs-mediated self interactions. This
is in the same class of models as our toy theory, although the DM particle $s$ could be fermionic in which case the spectrum of bound states and associated decays and transitions is even richer.   In all three cases the LHC search strategy for the DM state is 
greatly modified compared to the standard expectation of weak-scale neutralino DM.

\section{Conclusions}

We have shown that theories of TeV-scale physics can have dark matter candidates whose 
annihilation proceeds via the formation of near-threshold WIMPonium bound states.  Depending upon
the closeness-to-threshold of the weakest-bound state, these can lead to a substantial velocity-dependent
amplification of the dark matter annihilation cross-section,
preferring the lowest velocity dispersion environments all other factors being equal, and
providing a new dynamical source of ``boost factors" for indirect detection signals.  In addition,
the amplified radiative capture to more deeply bound states, and the transitions among such
bound states can both lead to a rich spectrum of discrete $\ga$ lines which, if observed, would
give striking confirmation of the mechanism.  

During the preparation of this work, \cite{posp} appeared. This paper also considers aspects of the phenomenology of WIMP bound states in the context of their annihilations and the consequences for dark matter indirect detection. 

\begin{acknowledgments}
We thank Markus Ahlers, Asimina Arvanitaki, Savas Dimopoulos, Peter Graham, Lawrence Hall, Roni Harnik, Dan Hooper, Karsten Jedamzik, Surjeet Rajendran, and Joe Silk for discussions. This work is partially supported by the EC Network 
6th Framework Programme Research and Training Network ``Quest for Unification"
(MRTN-CT-2004-503369) and by the EU FP6 Marie Curie Research and Training Network ``UniverseNet" (MPRN-CT-2006-035863). \end{acknowledgments}



\begin{thebibliography}{10}






  \bibitem{ams}
    J.~J.~Beatty {\it et al.},
  Phys.\ Rev.\ Lett.\  {\bf 93}, 241102 (2004)
  [arXiv:astro-ph/0412230];
  M.~Aguilar {\it et al.}  [AMS-01 Collaboration],
  Phys.\ Lett.\  B {\bf 646}, 145 (2007)
  [arXiv:astro-ph/0703154].

 
  \bibitem{atic}
  J.~Chang {\it et al.},
  Nature {\bf 456} (2008) 362.



 \bibitem{egret}
  D.~J.~Thompson, D.~L.~Bertsch and R.~H.~.~O'Neal,
  arXiv:astro-ph/0412376.
 
 \bibitem{glast}
  C.~Cecchi  [GLAST LAT Collaboration],
  J.\ Phys.\ Conf.\ Ser.\  {\bf 120} (2008) 062017.
 

 \bibitem{heat}
  S.~W.~Barwick {\it et al.}  [HEAT Collaboration],
  Astrophys.\ J.\  {\bf 482}, L191 (1997)
  [arXiv:astro-ph/9703192].

 \bibitem{hess}
  F.~Aharonian {\it et al.}  [HESS Collaboration],
  Astrophys.\ J.\  {\bf 636}, 777 (2006)
  [arXiv:astro-ph/0510397];
  F.~Aharonian  [HESS Collaboration],
  arXiv:0809.3894 [astro-ph].
  

 

 \bibitem{integral}
  A.~W.~Strong {\it et al.},
  Astron.\ Astrophys.\  {\bf 444}, 495 (2005)
  [arXiv:astro-ph/0509290];




 
 
 \bibitem{pamela}
  O.~Adriani {\it et al.},
  arXiv:0810.4995 [astro-ph];
  O.~Adriani {\it et al.},
  arXiv:0810.4994 [astro-ph].
 



  
  \bibitem{veritas}
  M.~Wood {\it et al.},
  arXiv:0801.1708 [astro-ph].
  
 
\bibitem{ourtalks}
Talk at PLANCK08, May 2008, JMR;
Seminars at the Universities of Southampton, Nottingham,  Cambridge and Oxford, Feb-Oct 2008 by SMW.

  
\bibitem{posp}
  M.~Pospelov and A.~Ritz,
  arXiv:0810.1502 [hep-ph].


  
 \bibitem{sommerfeld}
  A.~Sommerfeld, Ann. Phys. 11 257 (1931).
  

  
  \bibitem{freezeout}
  J.~Hisano, S.~Matsumoto and M.~M.~Nojiri,
  Phys.\ Rev.\  D {\bf 67}, 075014 (2003)
  [arXiv:hep-ph/0212022];
  J.~Hisano, S.~Matsumoto and M.~M.~Nojiri,
 Phys.\ Rev.\ Lett.\  {\bf 92}, 031303 (2004)
  [arXiv:hep-ph/0307216];
  J.~Hisano, \etal,
  Phys.\ Rev.\  D {\bf 71}, 015007 (2005)
  [arXiv:hep-ph/0407168];
  J.~Hisano, \etal,
  Phys.\ Rev.\  D {\bf 71}, 063528 (2005)
  [arXiv:hep-ph/0412403];
   S.~Profumo,
  Phys.\ Rev.\ D {\bf 72} (2005) 103521
  [arXiv:astro-ph/0508628];
  J.~Hisano, \etal
  Phys.\ Rev.\  D {\bf 73}, 055004 (2006)
  [arXiv:hep-ph/0511118];
 J.~Hisano, \etal,
  Phys.\ Lett.\  B {\bf 646} (2007) 34
  [arXiv:hep-ph/0610249];
    M.~Cirelli, A.~Strumia and M.~Tamburini,
  Nucl.\ Phys.\  B {\bf 787} (2007) 152
  arXiv:0706.4071 [hep-ph].




\bibitem{higgsportal}
  J.~March-Russell, S.~M.~West, D.~Cumberbatch and D.~Hooper,
  JHEP {\bf 0807} (2008) 058
  arXiv:0801.3440 [hep-ph].






\bibitem{others}
  C.~F.~Berger, L.~Covi, S.~Kraml and F.~Palorini,
  JCAP {\bf 0810}, 005 (2008)
  arXiv:0807.0211 [hep-ph];
  N.~Arkani-Hamed, D.~P.~Finkbeiner, T.~Slatyer and N.~Weiner,
  arXiv:0810.0713 [hep-ph];
  M.~Kamionkowski and S.~Profumo,
  arXiv:0810.3233 [astro-ph];
  L.~Ackerman, M.~R.~Buckley, S.~M.~Carroll and M.~Kamionkowski,
  arXiv:0810.5126 [hep-ph];
  I.~Cholis, D.~P.~Finkbeiner, L.~Goodenough and N.~Weiner,
  arXiv:0810.5344 [astro-ph];
    Y.~Nomura and J.~Thaler,
  arXiv:0810.5397 [hep-ph];
  D.~Feldman, Z.~Liu and P.~Nath,
  Phys.\ Rev.\  D {\bf 79}, 063509 (2009)
  [arXiv:0810.5762 [hep-ph]];
  Y.~Bai and Z.~Han,
  arXiv:0811.0387 [hep-ph];
  P.~J.~Fox and E.~Poppitz,
  arXiv:0811.0399 [hep-ph].




\bibitem{jomass}
  M.~Lattanzi and J.~Silk,
  arXiv:0812.0360 [astro-ph].



 
\bibitem{nocusp}
G.~Gentile, P.~Salucci, U.~Klein, D.~Vergani and P.~Kalberla,
  Mon.\ Not.\ Roy.\ Astron.\ Soc.\  {\bf 351} (2004) 903
  [arXiv:astro-ph/0403154].

\bibitem{subir}
  N.~W.~Evans, F.~Ferrer and S.~Sarkar,
  Phys.\ Rev.\  D {\bf 69}, 123501 (2004)
  [arXiv:astro-ph/0311145].



 \bibitem{bergetal}
  L.~Bergstrom and H.~Snellman,
  Phys.\ Rev.\  D {\bf 37}, 3737 (1988);
  S.~Rudaz,
  Phys.\ Rev.\  D {\bf 39}, 3549 (1989);
  A.~Bouquet, P.~Salati and J.~Silk,
  Phys.\ Rev.\  D {\bf 40}, 3168 (1989);
  G.~F.~Giudice and K.~Griest,
  Phys.\ Rev.\  D {\bf 40}, 2549 (1989);
  L.~Bergstrom,
  Phys.\ Lett.\  B {\bf 225}, 372 (1989);
  L.~Bergstrom,
  Nucl.\ Phys.\  B {\bf 325}, 647 (1989);
   L.~Bergstrom and P.~Ullio,
  Nucl.\ Phys.\  B {\bf 504} (1997) 27
  [arXiv:hep-ph/9706232];
    P.~Ullio and L.~Bergstrom,
  Phys.\ Rev.\  D {\bf 57} (1998) 1962
  [arXiv:hep-ph/9707333].
 
 \bibitem{bargmann}
V. Bargmann, {\it Proc.Nat.Acad.Sci. U.S.} {\bf 38} 961 (1952).
 

 
\bibitem{con}
C. Cohen-Tannoudji, B. Diu, F. Lalo\"{e}, Quantum Mechnics;
S.~De Leo and P.~Rotelli,
  Phys.\ Rev.\  D {\bf 78} (2008) 025006.
 
\bibitem{betheplaczek}
H.~A.~Bethe and G.~Placzek
Phys. Rev. 51, 450 - 484 (1937).

\bibitem{landl}
L.~D.~ Landau and E.~M.~ Lifshitz,
Quantum Mechanics, (Non-relativistic Theory), 3rd Ed, (New York: Pergamon).
 
 
 \bibitem{hulthepot}
 C.~Y.~Chen, F.~L.~Lu and D.~S.~Sun
 Phys.\ Scr.\ {\bf 76} (2007) 428-430.
 
 
 \bibitem{bethe}
 H.~A.~Bethe
 Phys. Rev. 47, 747 - 759 (1935).
 
 
 \bibitem{workinprogress}
  L.~J.~Hall, K.~Jedamzik, J.~March-Russell and S.~M.~West,
  arXiv:0911.1120 [hep-ph].
 

\bibitem{dwarfgalaxies}
  J.~D.~Simon and M.~Geha,
  Astrophys.\ J.\  {\bf 670} (2007) 313
  arXiv:0706.0516 [astro-ph].

 \bibitem{hiddenvalley}
  See for example M.~J.~Strassler and K.~M.~Zurek,
  Phys.\ Lett.\  B {\bf 651} (2007) 374
  [arXiv:hep-ph/0604261].

 

\end{thebibliography}
\end{document}